\documentclass[prl,preprint,showpacs,preprintnumbers,amsmath,amssymb]{revtex4}
\usepackage{graphicx}
\usepackage{dcolumn}
\usepackage{bm}
\usepackage[T1]{fontenc}
\usepackage{amsmath}

\begin{document}

\title{Temperature-dependent chirped coherent phonon dynamics in Bi$_{2}$Te$_{3}$ using high intensity femtosecond laser pulses}

\author{N. Kamaraju$ $,~Sunil Kumar$ $~and~A. K. Sood$ $\footnote{Electronic mail:~asood@physics.iisc.ernet.in}}

\affiliation{Center for Ultrafast Laser Applications (CULA)
and~$ $Department of Physics, Indian Institute of
Science,~Bangalore - 560 012, India }


\begin{abstract}
Degenerate pump-probe reflectivity experiments have been
performed on a single crystal of bismuth telluride
(Bi$_2$Te$_3$) as a function of sample temperature (3K to
296K) and pump intensity using $\sim$ 50 femtosecond laser
pulses with central photon energy of 1.57 eV. The time
resolved reflectivity data show two coherently generated
totally symmetric A$_{1g}$ modes at 1.85 THz and 3.6 THz at
296K which blue shift to 1.9 THz and 4.02 THz, respectively
at 3K. At high photoexcited carrier density of $\sim$ 1.7
$\times$ 10$^{21}$cm$^{-3}$, the phonon mode at 4.02 THz is
two orders of magnitude higher positively chirped (i.e the
phonon time period decreases with increasing delay time
between the pump and the probe pulses) than the lower
frequency mode at 1.9 THz. The chirp parameter, $\beta$ is
shown to be inversely varying with temperature. The time
evolution of these modes is studied using continuous
wavelet transform of the time-resolved reflectivity data.\\
\end{abstract}

\maketitle

\section{INTRODUCTION}
When a narrow band gap semiconductor is excited with
intense femtosecond laser pulses, a dense electron-hole
plasma is produced due to the promotion of electrons from
bonding states to anti-bonding states, which can cause
large changes in bond lengths leading to possible
structural transitions \cite{BiTe_Huang_PRL_1998,
BiTe_Siders_Science_1999}. The photo-excitation of carriers
changes the equilibrium positions of the atoms; the atoms
then oscillate around their new equilibrium positions, a
mechanism called as displacive excitation of coherent
phonons (DECP) \cite{BiTe_DECP_PRB_1992, BiTe_Cheng_APL,
BiTe_Kuznetsov}. Thus, DECP is the dominant mechanism in
opaque samples \cite{BiTe_DECP_PRB_1992} compared to
impulsive stimulated Raman scattering (ISRS)
\cite{BiTe_Nelson} in transparent materials. Later, it was
shown that DECP is a special case of ISRS when excited
resonantly \cite{BiTe_Garret_PRL_1996}. The investigations
of coherent phonons performed under high photo-excited
carrier density (PCD) $>$ 10$^{20}$ cm$^{-3}$ have been
very few till now \cite{BiTe_Hunsche_PRL_1995,
BiTe_Hunsche_APA, BiTe_Tangney_PRB_2002, BiTe_DeCamp_2001,
BiTe_M_Hase_PRL_2002, BiTe_Misochko_PRL_2004,
BiTe_Murray_2005}.

In tellurium, time resolved reflectivity experiments
\cite{BiTe_Hunsche_PRL_1995} performed using 100 fs pulses
at a PCD of $\sim 5 \times 10^{21}$ cm$^{-3}$ showed an
instantaneous large red shift (13\%) of $A_{1}$ coherent
phonon frequency, attributed to the electronic softening or
bond weakening \cite{BiTe_Stampfi_PRB_electron_Softening}.
In addition, the phonon time period decreasing with the
delay time between the pump and probe pulses corresponds to
an asymmetric line shape in frequency domain. This linear
sweep in the frequency with the pump-probe time delay,
termed as phonon chirping, originates from the rapid change
of photoexcited carrier density across the sample thickness
due to carrier diffusion resulting in different amounts of
phonon renormalization \cite{BiTe_Tangney_PRB_2002}. In
case of a semimetal, bismuth \cite{BiTe_M_Hase_PRL_2002,
BiTe_Misochko_PRL_2004} at carrier density of $\sim$ 3
$\times$ 10$^{21}$cm$^{-3}$ in a degenerate pump-probe
reflectivity experiment done at room temperature, A$_{1g}$
coherent phonons were also found to be positively chirped
with large oscillation amplitudes of $\sim$ 0.13 \AA
~\cite{BiTe_M_Hase_PRL_2002}. Further, collapse and revival
of chirped coherent phonon oscillations were observed in
the reflectivity data when the carrier density was
increased further beyond a critical carrier density levels
(3.5 $\times$ 10$^{21}$cm$^{-3}$ at 10K and 5 $\times$
10$^{21}$cm$^{-3}$ at 296K) \cite{BiTe_Misochko_PRL_2004}.
This behavior was explained in terms of dynamics of a
phonon wave packet in an anharmonic potential, where the
packet periodically breaks up and revives to its original
form, implying a nonclassical dynamics. A detailed first
principle density functional calculations and optical
double pump pulse experiments show that the anharmonic
contribution to the phonon period is negligible under such
high pump fluence regime in bismuth \cite{BiTe_Murray_2005}
and tellurium \cite{BiTe_Hunsche_PRL_1995,
BiTe_Tangney_PRB_2002} and time dependence of carrier
plasma density alone is able to explain the observed
softening of the phonons.

Now, we turn to Bi$_2$Te$_3$-subject matter of the present
study. Bi$_2$Te$_3$ is an important material both from the
point of view of thermoelectric \cite{BiTe_Nolas} as well
as exotic physics of topological insulators
\cite{BiTe_topological_Ins1,BiTe_topological_Ins2}. There
have been two earlier studies \cite{BiTe_Xu_APL_V92_2008,
BiTe_Xu_APL_V93_2008} of Bi$_2$Te$_3$ using non-degenerate
pump probe experiments at room temperature with 100 fs
pulses where pump fluence was kept below 1 mJ/cm$^2$. In
ref \cite{BiTe_Xu_APL_V92_2008}, the two observed coherent
phonon modes at 1.85 THz and 3.68 THz were assigned as
A$_{1g}$ and its second harmonic. In a later study
\cite{BiTe_Xu_APL_V93_2008}, the modes seen at 1.85 THz and
4.02 THz were assigned to two allowed A$_{1g}$ modes, in
agreement with conventional Raman measurements
\cite{BiTe_Richter_Raman_phys_stat_sol_1977}. The objective
of the present work is to study coherent phonons in single
crystal of Bi$_2$Te$_3$ as a function of temperature (from
296 K to 3 K) at high photoexcited carrier densities, with
a view (i) to understand the assignment of the high
frequency coherent mode (first order \emph{vis} a
\emph{vis} second order) and (ii) to study phonon chirping.
The Gabor wavelet transform has been performed on time
domain data to study the evolution of the coherent phonon
modes. We note that earlier studies on Bi$_2$Te$_3$
\cite{BiTe_Xu_APL_V92_2008, BiTe_Xu_APL_V93_2008} were done
only at room temperature and low pump fluences (with
negligible chirping).

\section{EXPERIMENTAL DETAILS}
A single crystal of Bi$_2$Te$_3$ (6 $\times$ 6 $\times$ 0.5
mm$^3$) with a cleaved surface perpendicular to the
trigonal axis mounted on a continuous helium flow cryostat
was used in our experiments. Femtosecond pulses were
derived from Ti:Sapphire amplifier (Spitfire, Spectra
Physics Inc) producing $\sim$ 50 fs pulses with the central
photon energy of 1.57 eV at a repetition rate of 1 KHz. The
pump beam was modulated at 393 Hz with a chopper and the
reflected probe intensity was recorded using a Si-PIN diode
and a lock-in amplifier. The spot size (half width at 1/e
maximum) of the pump and probe beams were kept at $\sim$
600 $\mu$m and 400 $\mu$m, respectively at the overlap of
the two beams on the sample. The pulse width was measured
to be 65~fs (full width at half maximum) using a thin
beta-barium borate (BBO) crystal at the sample point. The
polarization of the pump beam was kept perpendicular to
that of the probe beam to avoid scattered pump light
reaching the detector. Both the pump and probe beams were
kept close to normal incidence. The crystal surface is seen
to get damaged at pump fluence beyond $\sim$ 4.5 mJ/cm$^2$
and hence all our experiments were done at pump fluences of
3.3 mJ/cm$^2$ and 1.3 mJ/cm$^2$, whereas the probe fluence
was kept at 0.4 mJ/cm$^2$. The time resolved reflectivity
of the sample was recorded as a function of sample
temperature varying from 296K to 3K.

Bi$_2$Te$_3$ is a narrow band semiconductor with an
indirect band gap of 0.15 eV and it crystalizes in the
R$\overline{3}$m structure with the point group D$^5_{3d}$.
It is made up of close-packed atomic layers which are
periodically arranged along the c-axis in five layers
(Te$^{(1)}$-Bi-Te$^{(2)}$-Bi-Te$^{(1)}$) called as
`quintuples'. These layers are bonded by van der Waals
force and the weakest link among the layers is
Te$^{(1)}$-Te$^{(1)}$. Five atoms per hexagonal unit cell
(\emph{a} = 4.38~\AA,~ \emph{c} = 30.49~\AA) give totally
twelve optical phonons out of which four are Raman active
modes \cite{BiTe_Richter_Raman_phys_stat_sol_1977,
BiTe_E1_mode} represented as 2 A$_{1g} $(observed at 1.86
THz and 4.02 THz) + 2 E$_g$ (observed at 1.1 THz and 3.09
THz). The linear absorption coefficient, $\alpha$ of
Bi$_2$Te$_3$ at 1.57eV is $\sim$ 4 $\times$ 10$^5$
cm$^{-1}$ \cite{BiTe_Greenaway_JPCS_1965} and thus the
penetration depth, $\xi$ ($\sim 1/\alpha$) is 25 nm. This
corresponds to PCD, $N_0$ ($\equiv
6.25\times10^{18}~F\alpha(1-R)/E_p$) $\sim$ 1.7 $\times$
10$^{21}$ cm$^{-3}$ at pump fluence $F$ = 3.3 mJ/cm$^2$ at
the sample surface, which is about $1\%$ of all the valence
electrons. Here R=0.68 is the reflectivity coefficient of
Bi$_2$Te$_3$, at photon energy E$_p$ of 1.57 eV.

\section{RESULTS AND DISCUSSION}
\begin{figure}[htbp]
\centering\includegraphics[scale=0.68]{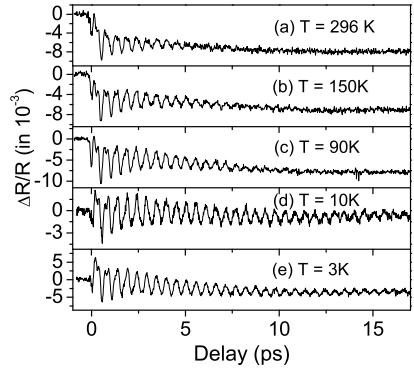}
\caption{Normalized time resolved reflectivity change
($\Delta$R/R) of bismuth telluride as the function of the
time delay between the pump and probe pulses at various
temperatures (a) T = 296 K, (b) 150 K , (c) 90 K, (d) 10 K,
(e) 3 K (colour online).}\label{BiTe_Fig1}
\end{figure}
\begin{figure}[htbp]
\centering\includegraphics[scale = 0.25]{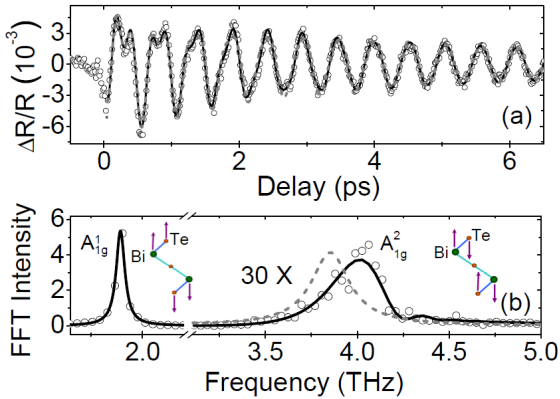}
\caption{(a) Digital band pass filtered normalized time
resolved differential reflectivity data at T = 3K (open
circles) along with their fits according to Eq.
\eqref{BiTe_DR_by_R1} (dashed line) and \eqref{BiTe_chirp}
(solid line). (b) The corresponding FFT of the time domain
data and the fits. The data in the range of 3 THz -5 THz
has been scaled appropriately to compare the asymmetry seen
at 3K. Here, the short dashed line is the FFT of the fit
without chirp parameter. The eigen vector of the two
A$_{1g}$ modes are shown in the FFT panel (colour
online).}\label{BiTe_Fig2}
\end{figure}

The coherent phonon mode's normal coordinate is written as
\cite{BiTe_A_V_Bragas_PRB_V69_205306_2004}, Q =
$b~exp(-\pi\gamma t)cos(2\pi \nu_0$t+$\phi$) where $b$,
$\gamma$, $\nu_0$ and $\phi$ are the amplitude, damping
constant, frequency and the initial phase of the coherent
phonon mode. To first order in Q, the normalized change in
reflectivity ($\Delta$R/R) of a probe beam due to
generation of two A$_{1g}$ coherent phonon modes
(A$^{(1)}_{1g}$ and A$^{(2)}_{1g}$) in an absorptive
material can be written as
\begin{equation}\label{BiTe_DR_by_R1}
 \frac{\Delta R}{R} = \sum_{i=1,2}\Bigg[\frac{\partial(\Delta R/R)}{\partial Q}\Bigg]~b_i
~exp(-\pi\gamma_it)~cos(2\pi \nu_it+\phi_i)
\end{equation}
Here $i=1$ and 2 corresponds to A$^{(1)}_{1g}$ and
A$^{(2)}_{1g}$ respectively. The time resolved reflectivity
data for Bi$_2$Te$_3$ at a few temperatures using pump
fluence of 3.3 mJ/cm$^2$ are shown in Fig. \ref{BiTe_Fig1}.
The signal contains both non-oscillatory and oscillatory
components. The non-oscillatory background arising from
carrier dynamics was removed by a digital band pass filter
to extract the oscillatory part \cite{BiTe_Xu_APL_V92_2008,
BiTe_Xu_APL_V93_2008}. The oscillatory part of the
transient normalized differential reflectivity data was
analyzed using Eq. \eqref{BiTe_DR_by_R1}. The fit was
satisfactory at room temperature but was not good at lower
temperatures. This necessitated the inclusion of chirp
parameter $\beta$ in Eq. \eqref{BiTe_DR_by_R1} as
\cite{BiTe_Misochko_PRL_2004}

\begin{equation}\label{BiTe_chirp}
    \frac {\Delta R} {R} = \displaystyle\sum_{i=1,2} B_{i}
    ~\mbox{exp}(-\pi\gamma_it)~cos(2 \pi \nu_i t + \beta_i t^2 + \phi_i)
\end{equation}
where $B_{i}=[\frac{\partial(\Delta R/R)}{\partial Q}]~b_i$
is the coherent phonon amplitude and $\beta_i$ is the chirp
parameter. This fit was found to be excellent over the
entire temperature range. For example, Fig.
\ref{BiTe_Fig2}(a) displays the digital band pass filtered
time domain data at 3K recorded using pump fluence of 3.3
mJ/cm$^2$ along with the fit using Eq.
\eqref{BiTe_DR_by_R1} (dashed line) and \eqref{BiTe_chirp}
(solid line), and Fig. \ref{BiTe_Fig2} (b) shows the
corresponding fast Fourier transform (FFT) of the data and
the fits. Here the FFT intensity of the second mode in the
frequency range of 3 THz - 5 THz is appropriately scaled up
to compare with the first mode. The eigen vectors
corresponding to the A$_{1g}$ modes
\cite{BiTe_Richter_Raman_phys_stat_sol_1977} are also shown
as the inset of Fig. \ref{BiTe_Fig2}(b). It may be very
difficult to see the chirping effect in time domain as the
higher frequency mode is very short lived (1.2 ps) and
therefore, frequency domain is a better choice in
identifying the chirp in such cases. The FFT of the fitted
function in time domain (Eq. \eqref{BiTe_chirp}) with chirp
($\beta \neq 0$) shown by continuous line) fits the FFT of
the measured data (open circles) much better than without
$\beta$, i.e $\beta = 0$ (short dashed line)[Fig.
\ref{BiTe_Fig2}(b)]. The value of $\beta_2 = 0.38$ps$^{-2}$
is much higher than $\beta_1$($\sim~2\times10^{-3}$
ps$^{-2}$). When the pump fluence is decreased to 1.3
mJ/cm$^2$, $\beta_2$ = 0.045 ps$^{-2}$ which clearly
indicates that the chirping is mainly due to large number
of photoexcited carriers. Since our focus is mainly on the
high density photo carriers mediated coherent phonons, the
temperature dependence of the fit parameters at highest
pump fluence of 3.3 mJ/cm$^2$ are discussed from here on.

\begin{figure}[htbp]
\centering\includegraphics[angle=0,scale=0.75]{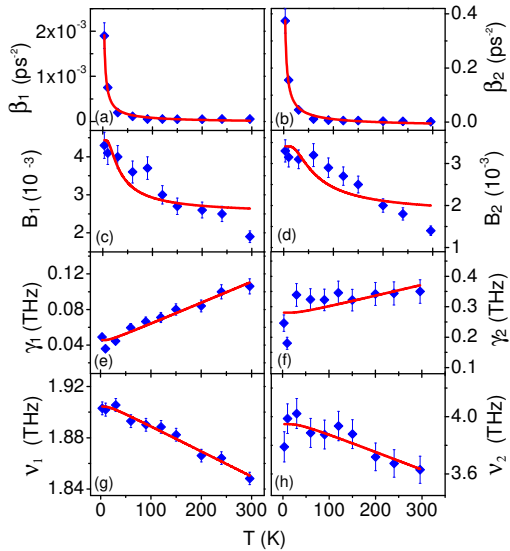}
\caption{Phonon chirp parameters [(a) and (b)], amplitude
of oscillations [(c) and (d)], damping constant [(e) and
(f)] and frequency [(g) and (h)] of both the coherent
A$_{1g}$ modes obtained from time domain fit to the data
with high pump intensity (Filled diamonds) versus sample
temperature. The continuous lines in (a)-(h) are the fits
(see text) (colour online).}\label{BiTe_Fig3}
\end{figure}

The parameters (filled diamond) obtained from the fit of
the time domain data are plotted as a function of
temperature in Fig. \ref{BiTe_Fig3}. Though the dependence
of $\beta_1$ and $\beta_2$ on temperature are similar, the
chirp parameter $\beta_2$ for the high frequency
A$^{(2)}_{1g}$ phonon mode is almost two orders higher
compared to $\beta_1$ showing that electron-phonon
interaction for A$^{(2)}_{1g}$ should be much higher than
that for A$^{(1)}_{1g}$(Figs. \ref{BiTe_Fig3}(a) and (b)).
The increased phonon chirping seen at low temperatures and
its inverse dependence on temperature $\beta \sim T^{-1}$
is shown using continuous lines. An increase of $\beta$
with decreasing temperature can be qualitatively understood
by considering the carrier diffusion across the penetration
depth of $\xi$ in Bi$_2$Te$_3$. At very high carrier
densities ($\sim$ 10$^{21}$), it has been shown that the
carrier diffusion coefficient, $D_a$ has major
contributions from carrier-carrier scattering mediated
diffusion, $D_{eh}$ \cite{BiTe_Fletcher, BiTe_C_M_Li_PRB}.
In such a case D$_a \sim T^{5/2}$ and hence the diffusion
time $\tau_{\textnormal{diff}} \equiv \xi^2/D_a \sim
T^{-5/2}$. This inverse power law dependence suggests that
at low temperatures, the diffusion time is longer and hence
electron phonon interaction will result in large chirping.
The temperature dependence of chirping as $\beta \sim
T^{-1}$ needs new theoretical inputs.

Next, we turn to the phonon amplitudes. It is seen that
both the coherent phonon amplitudes, B$_1$ and B$_2$
increase as the temperature is lowered (Figs.
\ref{BiTe_Fig3}.(c) and (d)). It has been shown
\cite{BiTe_Misochko_JPCM_2006} that the coherent phonon
amplitude, B$_{ph}$ behaves quite similar to the
temperature dependence of the Raman peak intensity
\cite{BiTe_Misochko_JPCM_2006} since the source of
spontaneous Raman scattering and the driving force in the
generation of coherent phonons through ISRS are the same
\cite{BiTe_Merlin_SSC}. The Raman cross-section increases
with temperature as [$n(\nu)+1$] (where n($\nu$) is the
Bose-Einstein statistical factor) and hence the Raman peak
intensity can be written as, I$_p$ $\sim
\frac{[n(\nu)+1]}{[2 n(\nu/2)+1]}$, where the temperature
dependence in the denominator comes from the cubic
anharmonic contributions to the linewidth. The fit (solid
line) using this expression is shown along with the data
for B$_{1}$ and B$_{2}$ in Fig. \ref{BiTe_Fig3}(c) and (d).

The temperature dependence of coherent phonon damping
factor and the frequency of the two modes are displayed in
Fig. \ref{BiTe_Fig3}(e),(g) and Fig.
\ref{BiTe_Fig3}(f),(h), respectively. In the case of
A$^{(1)}_{1g}$, the frequency decreases from 1.9 THz to
1.84 THz, i.e., a decrease of 3$\%$ as the crystal is
heated from 3K to 296K and the damping term increases by
150 $\%$ from $\sim$ 0.04 THz at 3K to $\sim$ 0.1 THz at
296K. The behavior of the second mode is rather
interesting: the damping constant becomes almost constant
after 60K and the frequency decreases from 4.0 THz to 3.6
THz (10$\%$ change) when the crystal temperature increases
from 3K to 296K. The fit (thick line) shown in the figure
is by using the well known functions \cite{BiTe_Balkanski}
based on cubic anharmonicity where the phonon of frequency,
$\nu$ decays into two phonons of equal frequency:
$\gamma_{ph}(T) = \gamma_0 + C[1 + 2n(\nu_0/2)]$ and
$\nu_{ph}(T)= \nu_0 + A [1 + 2n(\nu_0/2)]$ where $\nu_0$
(frequency at T=0K), A, C and $\gamma_0$ (disorder induced
damping) are the fitting parameters (A and C are the
measures of third order cubic anharmonicity). The
parameters obtained from fitting are, $\nu_0$ = 1.908 THz,
A = -0.005 THz, $\gamma_0$ = 0.040 THz and C = 0.005 THz
for A$^{(1)}_{1g}$ mode; and $\nu_0$ = 4.007 THz, A =
-0.060 THz, $\gamma_0$ = 0.263 THz and C = 0.017 THz for
A$^{(2)}_{1g}$ mode. Thus the disorder induced damping
($\gamma_0$) and anharmonicity (A and C) are more for
A$^{(2)}_{1g}$ compared to A$^{(1)}_{1g}$. It can be seen
from Fig. \ref{BiTe_Fig3}(h) that the temperature
dependence of $\nu_2$ is non-monotonic. The decrease of
$\nu_2$ at 3K may be due to the chirp in the frequency
whose quantitative understanding is lacking.

\begin{figure}[htbp]
\centering\includegraphics[width= 7.5 cm]{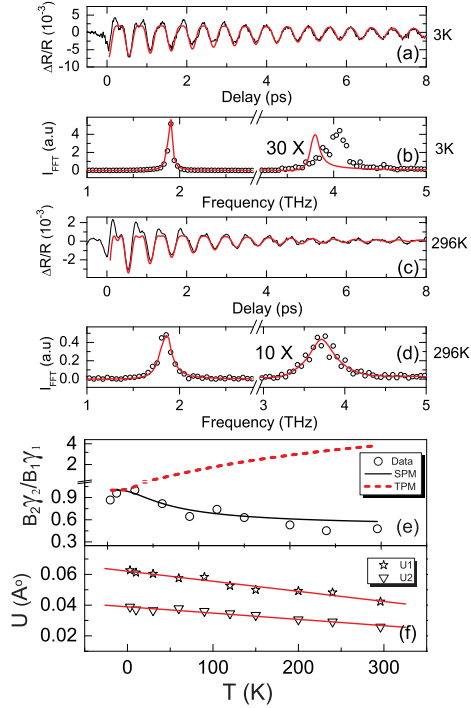}
\caption{The time domain data (thin line) along with the
fit (thick line) according to Eq. \eqref{BiTe_DR_by_R2} and
their FFT are shown in (a)-(b) for T=3K and (c)-(d) for
T=296K. In (b) and (d), the open circles are the FFT of the
data and thick line is the FFT of the fit. The ratio of
integrated coherent phonon intensity (open circles),
$\frac{B_2\gamma_2}{B_1\gamma_1}$ along with the fit using
single phonon (SPM) (thick line) and two phonon model (TPM)
(dashed line) are shown in (e). The estimated lattice
displacements for the two coherent phonon modes (open star
for U$_1$ and open inverted triangle for U$_2$) are given
in (f) along with a linear fit (colour online).}
\label{BiTe_Fig4}
\end{figure}

Finally, to examine whether the second mode at $\sim$ 4 THz
is the second harmonic of the first A$_{1g}$ mode at 1.8
THz as suggested in ref \cite{BiTe_Xu_APL_V92_2008}, we
re-fitted the digitally band pass filtered time domain data
with $\frac{\Delta R}{R}$ expressed in a second order
approximation in Q as \cite{BiTe_Xu_APL_V92_2008}
\begin{equation}\label{BiTe_DR_by_R2}
 \frac{\Delta R}{R} = \Bigg[\frac{\partial(\Delta R/R)}{\partial
 Q}\Bigg]~Q + \frac{1}{2}\Bigg[\frac{\partial^2(\Delta R/R)}{\partial
 Q^2}\Bigg]~Q^2
\end{equation}
where Q is for the A$^{(1)}_{1g}$ phonon mode. Taking the
expression for Q as in Eq. \eqref{BiTe_DR_by_R1} (since
$\beta_1$ is small), the fit according to Eq.
(\ref{BiTe_DR_by_R2}) was unsatisfactory as the second mode
could never be accommodated in the fit to the data as shown
in Fig. \ref{BiTe_Fig4}. In Fig. \ref{BiTe_Fig4}, the time
domain data (thin line) along with the new fit (thick line)
using Eq. \eqref{BiTe_DR_by_R2} and their FFT are shown in
(a)-(b) for 3K and in (c)-(d) for 296K where open circles
are the FFT of the data and thick lines are the FFT of the
fit. It can be seen from these figures that the fit is not
compatible with the data in both frequency and time domain.
At 296K, though the fit seemingly agrees with the data in
the frequency domain, it is not so in time domain. This
confirms that the higher energy phonon mode is not a second
order mode of A$^{(1)}_{1g}$. To further elucidate this
point, we have analyzed the temperature dependence of the
ratio of the integrated intensities of coherent phonons,
$\frac{B_2\gamma_2}{B_1\gamma_1}$ (open circles in
Fig.\ref{BiTe_Fig4}(e) where the y-axis is normalized to
the ratio at 3K). Here, we consider two cases: (i) For
$\nu_1$ and $\nu_2$ with $\nu_2 > \nu_1$ as the single
phonon modes (SPM), the ratio of their integrated
intensity, is $\sim$ $\frac{n(\nu_2)+1}{n(\nu_1)+1}$ shown
by solid line in Fig. \ref{BiTe_Fig4}(e), (ii) for $\nu_2$
as the second harmonic or two phonon mode (TPM) of $\nu_1$
($\nu_2= 2 \nu_1$), we note that the ratio of their
integrated intensity is $[n(\nu_1)+1]^2/[n(\nu_1)+1]$ =
$n(\nu_1)+1$ shown by dashed line in Fig. \ref{BiTe_Fig4}
(e). Thus, it is clear from Fig. \ref{BiTe_Fig4} (e) that
SPM is compatible with the data corroborating our
conclusion that the high frequency mode is a single phonon
mode and not a second harmonic of A$^{(1)}_{1g}$.

The lattice displacement of the coherent phonon modes at 1.9 THz
and 4.02 THz can be estimated for absorbing materials using
\cite{BiTe_DeCamp_2001, BiTe_Decamp_Thesis}
\begin{equation}\label{BiTe_U1}
    U_i^2 \sim \frac{3.8 \times 10^{-3}~B_i~F}{\varrho \nu_i |\varepsilon|}
\left[\frac{(\frac{2\varepsilon_2}{E_{ph}})}{D}\right]
\end{equation}
where $U_i$ is in Angstrom (\AA), $F$ is the pump fluence in
mJ/cm$^2$, $\varrho$ is the density of the material in amu/\AA$^3$
and $\varepsilon$ is the dielectric constant
($\varepsilon_1+j\varepsilon_2$), E$_{ph}$ is the energy of the
phonon in eV and $D = \frac{1}{R}\frac{\partial R}{\partial E}$
with E as the photon energy in eV. Now, for Bi$_2$Te$_3$, $D =
\frac{1}{R}\frac{\partial R}{\partial E} \sim 10^{-1} eV^{-1}$ at
1.57 eV and $\frac{2\varepsilon_2}{E_{ph}} \sim 10^3 eV^{-1}$ with
$\varepsilon$ = 2.75 + $j$ 15 at 1.57 eV
\cite{BiTe_Greenaway_JPCS_1965}. Thus, for Bi$_2$Te$_3$, the
lattice displacement is
\begin{equation}\label{BiTe_equation5}
U_i \sim \sqrt{B_i \frac{38~F}{\varrho \nu_i
|\varepsilon|}}
\end{equation}
The temperature dependence of U$_1$ and U$_2$ thus
estimated is shown in Fig. \ref{BiTe_Fig4} (f), which
essentially arises from temperature dependence of B$_i$ and
$\nu_i$. A linear fit (line) to U is shown in the figure.

To capture the evolution of coherent phonons with time, we
have performed continuous wavelet transform (CWT) similar
to that used by Hase et al. \cite{BiTe_M_Hase_Science} in
observing the birth of a quasiparticle in silicon. We have
used the MATLAB code \cite{BiTe_coloroda} modified for
Gabor mother wavelet based on Gaussian function given as
\cite{BiTe_Suzuki_JAE_V14_p69_1996},
\begin{equation}\label{BiTe_gabor}
    \Psi(t/s) =
    \pi^{-1/4}\left(\frac{1}{ps}\right)^{1/2}~exp\left[-\frac{t^2}{2s^2p^2} + j \frac{t}{s}\right]
\end{equation}
where $s$ is the scaling factor (inverse of frequency) and
$p$=$\pi(2/ln(2))^{1/2}$ is a constant. Here, we describe
the procedure to calculate the continuous wavelet
transform. The wavelet transform of a given time signal,
$x$(t) is given by,
\begin{eqnarray}\label{BiTe_CWT}
    CWT(\tau,s) & = & \frac{1}{\sqrt{|s|}}
                 \int x(t)\Psi^*((t-\tau)/s) dt
\end{eqnarray}
\begin{figure}[htbp]
\centering\includegraphics[scale=0.3]{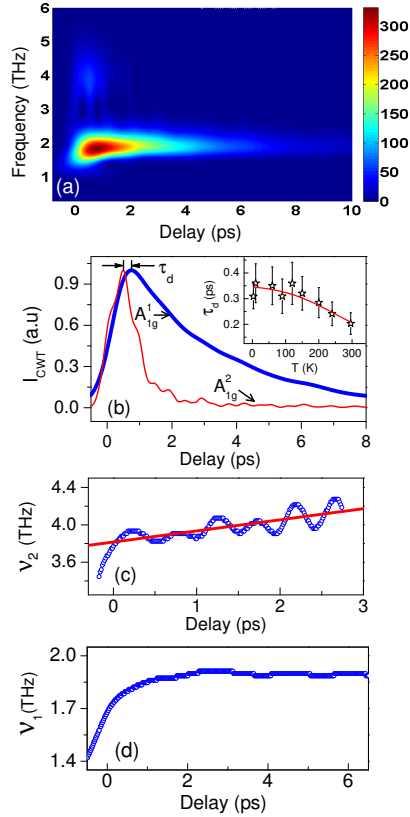}
\caption{(a) CWT chronogram of high pump intensity data at
T=3K. The color bars are given at the right side to
symbolically represent the CWT intensity. (b) The maximum
of CWT intensity at each time delay for both modes (thicker
line is for A$^{(1)}_{1g}$ mode and the thinner line is for
the A$^{(2)}_{1g}$ mode). The time delay between the two
modes (open stars) are plotted as a function of sample
temperature in the inset where the thick line is the guide
to the eye. (c) The frequency corresponding to the maxima
of CWT intensity at each time delay for A$^{(2)}_{1g}$
(thicker line is the linear fit to the data) and for (d)
A$^{(1)}_{1g}$ mode. (see text) (color
online).}\label{BiTe_Fig5}
\end{figure}
A starting scale, $s_{start}$ corresponding to a frequency
higher than the highest frequency of the signal (determined
by FFT) is chosen and the starting wavelet
$\Psi_{start}(t/s_{start})$ in time domain here is a
compressed wavelet. The cross correlation of
$\Psi_{start}(t)$ with $x(t)$ is computed using Eq.
\eqref{BiTe_CWT} with $\tau = 0$. The magnitude of cross
correlation will depend on how closely the frequency
components in $x(t)$ and $\Psi_{start}(t)$ match. This
procedure is repeated by translating $\Psi_{start}(t)$ in
time domain ($\tau$) and this gives wavelet coefficients
for a given $s$ and a range of values of $\tau$. The whole
procedure is again repeated for next higher $s$. Thus, one
gets a range of wavelet coefficients of $x(t)$ in
time-scale plane which is then converted into
time-frequency plane. The wavelet transform of a time
domain signal gives three dimensional (3D) plot of wavelet
coefficients vs frequency and time. CWT chronograms
(contour of the 3D plot) of the time domain data at 3K and
high pump excitation (3.3 mJ/cm$^2$), are shown in Fig
\ref{BiTe_Fig5} (a). It can be seen from Fig
\ref{BiTe_Fig5} (a) that at 3K, the frequency of
A$^{(1)}_{1g}$ mode starts at 1.5 THz and reaches 1.9 THz
in $\sim$ 750 fs and this time is seen to be constant at
all the temperatures. We could not resolve a similar build
up in frequency for the A$^{(2)}_{1g}$ due to its short
lifetime (0.9-1.6 ps) compared to the first mode (4-10 ps).
To get more insight into these chronograms, the maximum of
CWT intensity at each time delay (time slice) was
calculated and plotted at 3K for both the modes (Fig.
\ref{BiTe_Fig5}(b)). Here, thicker line is for
A$^{(1)}_{1g}$ and the thinner line is for A$^{(2)}_{2g}$.
The time to reach the maximum of CWT is $\sim$750 $\pm$ 20
fs fs for A$^{1}_{1g}$ mode and $\sim$ 510 $\pm$ 20 fs for
A$^{2}_{1g}$ mode. The time delay of $\sim$ 240 $\pm$ 20 fs
between the maxima of the two modes A$^{1}_{1g}$ and
A$^{2}_{1g}$ is found to dependent on the sample
temperature as shown in the inset of Fig. \ref{BiTe_Fig5}
(b). The reason for the time delay and its temperature
dependence is yet to be understood. The phonon chirp for
the second mode is demonstrated through the plot of
frequency(open circles) corresponding to the maxima of CWT
intensity at each time delay at 3K in Fig. \ref{BiTe_Fig5}
(c). Thicker line is the linear fit to the data with $\nu_2
(THz) = 3.8 + \beta_2/(2\pi)t$ where t is the time delay in
ps. This corresponds to $\beta_2 = 0.41$ ps$^{-2}$, closely
matching to the value of 0.38 ps$^{-2}$ derived from the
time domain fit. The oscillatory structures seen in the
frequency is an artifact in the wavelet transform when the
separation between the two frequencies is less than 4 THz.
Similar analysis for A$^{1}_{1g}$ mode is given in Fig.
\ref{BiTe_Fig5} (d) where the build-up time is $\sim$ 750
fs (consistent with Fig. \ref{BiTe_Fig5} (b)) and the
frequency is seen to be constant with delay time as
expected since $\beta_1$ is very small. We note that the
above analysis using AGU-Vallen-wavelet \cite{BiTe_Vallen}
resulted in similar results.

\section{CONCLUSIONS}
In conclusion, we have studied coherent A$_{1g}$ phonons in
Bi$_2$Te$_3$ as a function of both temperature and pump
fluence. We have observed that the higher frequency
coherent phonon mode at $\sim$ 4.0 THz is A$^{(2)}_{1g}$
and is not a second harmonic of A$^{(1)}_{1g}$. It acquires
two orders of magnitude higher positive chirping at the
lowest temperatures and high pump fluence as compared to
the A$^{(1)}_{1g}$ phonon. The chirp in time domain is
manifested as an asymmetry in the frequency domain. The
wavelet transform of the time domain differential
reflectivity helps to identify the chirp and the time for a
phonon mode to build up to its maximum amplitude. We hope
our experiments will motivate theoretical calculations to
understand the exact temperature dependence of phonon
chirping at highly excited carrier densities, higher
chirping for A$^{(2)}_{1g}$ and the different build up
times for two coherent modes.

\acknowledgments AKS acknowledges the financial support from
Department of Science and Technology of India and CSIR for the
Bhatnagar Fellowship. SK acknowledges University Grants
Commission, India for senior research fellowship.


\begin{thebibliography}{10}
\providecommand*{\bibinfo}[2]{#2}
\providecommand*{\eprint}[1]{#1}
\providecommand*{\url}[1]{#1}

\bibitem{BiTe_Huang_PRL_1998}
L. Huang, J. P. Callan, E. N. Glezer, and E. Mazur, Phys.
Rev. Lett. \textbf{80}, 185 (1998).

\bibitem{BiTe_Siders_Science_1999}
C. W. Siders, A. Cavalleri, K. S-Tinten,Cs. T$\acute{o}$th,
T. Guo, M. Kammler,5 M. H. v. Hoegen, K. R. Wilson, D. von
der Linde, C. P. J. Barty, Science 286, 1340 (1999).

\bibitem{BiTe_DECP_PRB_1992}
H. J. Zeiger, J. Vidal, T. K. Cheng, E. P. Ippen, G.
Dresselhaus, M. S. Dresselhaus, Phys. Rev. B. \textbf{45}
768 (1992).

\bibitem{BiTe_Cheng_APL}
T. K. Cheng, M. S. Dresselhause, and E. P. Ippen, Appl.
Phys. Lett. \textbf{62}, 1901 (1993).

\bibitem{BiTe_Kuznetsov}
A. V. Kuznetsov and C. J. Stanton, Phys. Rev. Lett.
\textbf{73}, 3243 (1994).

\bibitem{BiTe_Nelson}
L. Dhar, J. A. Rogers and K. A. Nelson, Chem. Rev.
\textbf{94}, 157 (1994).

\bibitem{BiTe_Garret_PRL_1996}
G. A. Garret, T. F. Albrecht, J. F. Whitaker, and R.
Merlin, Phys. Rev. Lett. \textbf{77}, 3661 (1996).

\bibitem{BiTe_Hunsche_PRL_1995}
S. Hunsche, K. Wienecke, T. Dekorsy, and H. Kurz, Phys.
Rev. Lett. \textbf{75}, 1815 (1995).

\bibitem{BiTe_Hunsche_APA}
S. Hunsche,K. Wienecke, and H. Kurz, Appl. Phys. A
\textbf{62}, 499 (1996).

\bibitem{BiTe_Tangney_PRB_2002}
P. Tangney, S. Fahy, Phys. Rev. B \textbf{65}, 054302
(2002).

\bibitem{BiTe_DeCamp_2001}
M. F. DeCamp, D. A. Reis, P. H. Bucksbaum, and R. Merlin,
Phys. Rev. B \textbf{64}, 092301 (2001).

\bibitem{BiTe_M_Hase_PRL_2002}
M. Hase , M. Kitajima, S. Nakashima, and M. Mizoguchi,
Phys. Rev. Lett. \textbf{88}, 067401 (2002).

\bibitem{BiTe_Misochko_PRL_2004}
O. V. Misochko , M. Hase, K. Ishioka, and M. Kitajima,
Phys. Rev. Lett. \textbf{92}, 197401 (2004).

\bibitem{BiTe_Murray_2005}
E. D. Murray, D. M. Fritz, J. K. Wahlstrand, S. Fahy, and
D. A. Reis, Phys. Rev. B \textbf{72}, 060301 (2005).

\bibitem{BiTe_Stampfi_PRB_electron_Softening}
P. Stampfli and K. H. Bennemann, Phys. Rev. B \textbf{42},
7163 (1990); \textbf{46}, 10686 (1992); \textbf{49}, 7299
(1994).

\bibitem{BiTe_Nolas}
G. S. Nolas, J. Sharp, and H. J. Goldsmid,
\textbf{Thermoelectrics} (Springer: New York, 2001)

\bibitem{BiTe_topological_Ins1}
C. L. Kane and E. J. Mele, Science, \textbf{314}, 1692
(2006).

\bibitem{BiTe_topological_Ins2}
J. E. Moore, Nature \textbf{464}, 194 (March 2010).

\bibitem{BiTe_Xu_APL_V92_2008}
A. Q. Wu, X. Xu, and R. Venkatasubramanian, Appl. Phys.
Lett. \textbf{92}, 011108 (2008).

\bibitem{BiTe_Xu_APL_V93_2008}
Y. Wang, X. Xu, and R. Venkatasubramanian, Appl. Phys.
Lett. \textbf{93}, 113114 (2008).

\bibitem{BiTe_Richter_Raman_phys_stat_sol_1977}
W. Richter, H. Kohler, and C. R. Becker, Phys. Stat. Sol. (b)
\textbf{84}, 619 (1977).

\bibitem{BiTe_E1_mode}
W. Kullmann, J. Geurts, W. Richter, N. Lehner, H. Rauh, U.
Steigenberger, G. Eichhorn, and R. Geick, Phys. Stat. Sol.
(b) \textbf{125}, 131 (1984).

\bibitem{BiTe_Greenaway_JPCS_1965}
D. L. Greenaway and G. Harbeke, J. Phys. Chem. Solids
\textbf{26}, 1585 (1965).

\bibitem{BiTe_A_V_Bragas_PRB_V69_205306_2004}
A. V. Bragas, C. Aku-Leh, S. Costantino, A. Ingale, J. Zhao
and R. Merlin, Phys. Rev. B \textbf{69}, 205306 (2004).

\bibitem{BiTe_Fletcher}
N. H. Fletcher, Proc. Inst. Rad. Eng. \textbf{45}, 862
(1957).

\bibitem{BiTe_C_M_Li_PRB}
C. -M. Li, T. Sjodin, and H. -L. Dai, Phys. Rev. B
\textbf{56}, 15 252 (1997).

\bibitem{BiTe_Misochko_JPCM_2006}
O. V. Misochko, K. Ishioka, M. Hase
and M. Kitajima, J. Phys.:Condens. Matter \textbf{18},
10571 (2006).

\bibitem{BiTe_Merlin_SSC}
R. Merlin, Solid State Commun. \textbf{102}, 207 (1997).

\bibitem{BiTe_Balkanski} M. Balkanski, R. F. Wallis, E. Haro,
Phys. Rev. B 28, 1928 (1983).

\bibitem{BiTe_Decamp_Thesis}
See Eq. A.22 to A.24 of M. F. Decamp, Ph.D Thesis, The
University of Michigan, Ann Arbor (2002), available at
http://www.aps.anl.gov/Sectors/Sector7/Science/
Publications/theses.html. Note that in Eq. A.22, there is a
typographical error in the exponent of
($\frac{2i\varepsilon_2}{E_{ph}}$) which should be 1
instead of 2.

\bibitem{BiTe_M_Hase_Science}
M. Hase, M. Kitajima, A. M. Constantinescu, and H. Petek,
Nature \textbf{426}, 51 (2003).

\bibitem{BiTe_coloroda}
Wavelet software, http://paos.colorado.edu/research
/wavelets/software.html

\bibitem{BiTe_Suzuki_JAE_V14_p69_1996}
H. Suzuki, T. Kinjo, Y. Hayashi, M. Takemoto, K. Ono, J.
Acoustic Emission \textbf{14}, 69 (1996).

\bibitem{BiTe_Vallen}
AGU-Vallen Wavelet, http://www.vallen.de/wavelet
/index.html

\end{thebibliography}
\end{document}